# Mid-infrared optics using dielectrics with refractive indices below unity

Alireza Shahsafi[1], Yuzhe Xiao[1], Jad Salman[1], Bradley S. Gundlach[1], Chenghao Wan[1,2], Patrick J. Roney[1], Mikhail A. Kats[1,2,3]

Departments of [1]Electrical and Computer Engineering, [2]Materials Science and Engineering, and [3]Physics

University of Wisconsin-Madison

**Abstract**

Conventional transparent materials at optical frequencies have refractive index values (*n*) greater than unity – most commonly between about 1 and 4. This paper explores optical phenomena made possible by using materials with refractive indices less than unity. We focus primarily on fused silica ($SiO_2$), a polar dielectric with strong phonon resonances in the mid infrared that result in a spectral region in which $n < 1$ with modest optical loss. Using this ubiquitous easy-to-deposit material, we demonstrate infrared-frequency external reflection, frustrated external reflection, and direct coupling to surface plasmon polaritons from free space. Our work suggests that materials with refractive indices below unity can bring significant new functionality to optical devices.

# 1. Introduction

Traditional refractive optical components (*e.g.,* lenses) are made from materials that are transparent in the spectral regions for which they are designed, and typically have refractive indices ($n$) between 1 and 4 [1]. In such regions, these materials are devoid of optical resonances, resulting in negligible optical losses. Recently, the use of both naturally occurring and engineered resonances has enabled the demonstration of a number of exotic optical phenomena, especially in the field of metamaterials [2]. For example, resonances in both the permittivity, $\epsilon$ and permeability, $\mu$, can result in a negative index of refraction ($n = \sqrt{\epsilon\mu} < 0$) [3]–[5]. Similarly, by using electric or magnetic resonances, the values of $\epsilon$ or $\mu$ can be made to approach zero (*i.e.,* "epsilon-near-zero", or "mu-near-zero", respectively), resulting in a near-zero refractive index ($n \sim 0$) [6], [7]. In such a material, light can propagate without accumulating phase, enabling unique phenomena such as super-coupling [8]. Unfortunately, in many instances, the promise of these exotic phenomena and applications is tempered by optical losses, which are enhanced in close proximity to resonances [9], [10].

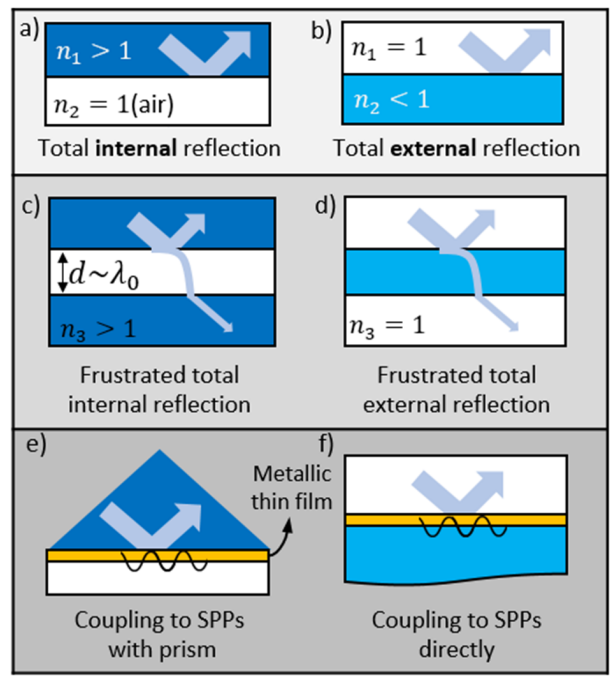

**Figure 1)** Three well-known optical phenomena (left), and their less-conventional counterparts enabled by materials with $n < 1$ (right). **(a)** Total reflection for light incident from a high-index region (1) to air (2), and **(b)** from air (1) to a low-index region (2). **(c-d)** If the thickness of region 2 is comparable to the wavelength of light, $\lambda_0$, and a higher-index region 3 is introduced below, some light can be transmitted. This is referred to as frustrated total internal reflection if $n_1 > 1$; by analogy, we refer to the case when $n_1 = 1$ as frustrated total *external* reflection. **(e)** Conventional prism coupling to surface plasmon polaritons (SPPs), known as the Kreschmann configuration. **(f)** Coupling to SPPs on the interface between metal and a material with $n < 1$ can be achieved with no prism or surface modulation.

In this work, we explore optical phenomena within a spectral range that is close — but not too close — to strong optically active resonances, resulting in the reduction of the real part of the refractive index to below that of vacuum (*i.e.*, $0 < n < 1$). Our demonstrations use a simple homogeneous material, fused silica ($SiO_2$), which has strong optical-phonon resonances in the mid infrared [11], but is transparent in the visible and near infrared. The intermediate proximity to these optical-phonon resonances results in only a modest increase in the losses, quantified by a small imaginary part of the refractive index, $\kappa$. By utilizing the $n < 1$ region in $SiO_2$, we experimentally demonstrate several unusual optical phenomena in the mid infrared, including external reflection (Fig. 1b) and its frustration (Fig. 1d), and direct coupling to surface-plasmon polaritons without the use of prisms or surface modulation (Fig. 1f).

**Optical properties of $SiO_2$**

In polar dielectric materials — *e.g.,* $SiO_2$ or aluminum oxide ($Al_2O_3$) — optical-phonon resonances typically lie within the mid-infrared spectral range. The resulting optical properties can be described by Lorentz [12], Gaussian [13], or similar oscillator models and, in the vicinity of such resonances, the refractive index can be below unity [11].

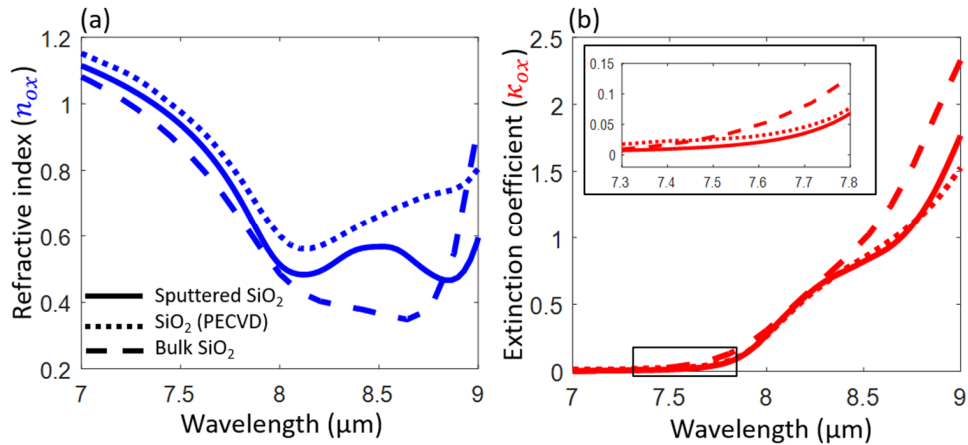

**Figure 2)** Optical properties of $SiO_2$. **(a)** Real ($n_{ox}$) and **(b)** imaginary ($\kappa_{ox}$) parts of the complex refractive index of fused silica ($SiO_2$) close to its phonon resonances. Two sets of experimentally obtained refractive indices are presented: one of a film grown via plasma-enhanced chemical vapor deposition (PECVD, dotted), and the other from an $SiO_2$ wafer. For comparison, data taken from ref. [11] are shown using solid lines. Inset: zoomed-in plot of $\kappa$ in a region where $n_{ox} < 1$ and the losses are relatively small.

In this study, we used two types of $SiO_2$ for the various experiments: thin films grown via plasma-enhanced chemical vapor deposition (PECVD) on high-resistivity silicon, and bulk $SiO_2$ wafers. The measured refractive index, $n_{ox}$, and extinction coefficient, $\kappa_{ox}$, of these are shown in the Fig. 2 (see Supplementary Information 1 for more details), and compared to literature data [11]. Though the optical properties differ somewhat between the different samples, in all cases $\kappa_{ox}$ is relatively small as $n_{ox}$ approaches 1. Specifically, in the 7.3 ~ 7.7 μm region, $n_{ox} < 1$ and $\kappa_{ox} < 0.05$.

## 2. Results
### 2.1. Greater reflection than noble metals

When light is incident on a low-index ($n_2$) material from a high-index ($n_1$) material, total internal reflection (TIR) occurs for incident angles greater than the critical angle ($\theta_{cr} = \arcsin(n_2/n_1)$) (Fig. 1a). In TIR, the incident medium is usually an optically dense material such as glass. Analogously, when the incident medium is air, and the second material has $n < 1$, a similar phenomenon occurs, referred to as total external reflection (TER). TER is well-known at X-ray frequencies, where complex refractive indices are often written in the form: $\tilde{n} = 1 - \Delta + i\beta$, where typically $\Delta \sim 10^{-5}$ and $\beta \sim 10^{-7}$ [14]. Mirrors based on TER are used for X-ray focusing and other applications [15], [16].

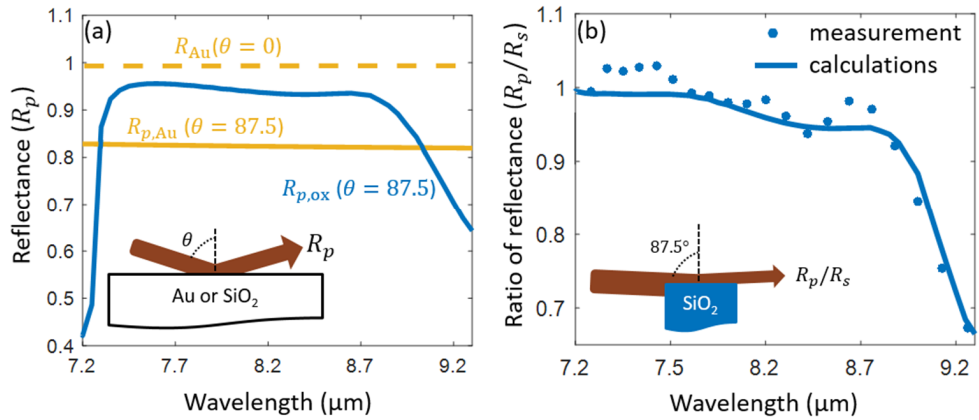

**Figure 3)** Oblique-angle reflection of SiO$_2$ can surpass that of gold. **(a)** Calculated p-polarized reflectance spectra for gold (dashed and solid yellow for normal and oblique incidence, respectively) and an SiO$_2$ wafer (blue) for incident light coming from air. At large incident angles ($\theta = 87.5°$), the p-polarized reflectance of gold and other metals is suppressed due to the pseudo-Brewster effect. In the wavelength range in which SiO$_2$ has lower refractive index than unity, the high-angle reflectance of SiO$_2$ is above that of gold due to external reflection. **(b)** Calculated (solid line) and experimental (circles) ratio of the p- and s-polarized reflectance spectra for our SiO$_2$ wafer at $\theta = 87.5°$.

In the mid infrared, metallic mirrors can have reflectances of ~ 99% [17], and are often used as references for reflectance measurements. However, for very oblique incidence angles and p polarization, even gold and silver mirrors have a reduced reflectance due to the pseudo-Brewster effect [18], [19] (more discussion in Supplementary Information 2). Polished gold mirrors, for example, have significantly decreased reflectance at angles of incidence between 80° and 89° in the near to mid infrared (*e.g.*, Fig. 3a). By comparison, due to TER, materials with $n < 1$ can be used to achieve much larger p-polarized reflectance for these highly oblique angles. We calculated and measured the p-polarized reflectance of an $SiO_2$ wafer at an incident angle of 87.5° (Fig. 3), and found reflectance values of ~ 95%, compared to ~ 82% for gold for $\lambda_0$ =7.3 ~ 8.7 $\mu m$; the $SiO_2$ p-polarized reflectance is higher than that of gold for angles > 85° (more Supplementary Information 2). We note that due to the highly oblique angle of incidence and relatively small sample area, we were not able to accurately measure the absolute reflectance, because the beam cross section was larger than our sample (see Fig. 3b inset). Instead, we measured the ratio of p-polarized and s-polarized reflectance, eliminating the need for a reference (Fig. 3b). This figure shows good agreement between the measurement and calculations based on Fresnel equations, using the refractive index data in Fig. 3a.

### 2.2. Frustrated external reflection

In both TIR and TER, light is incident on a material with a lower refractive index from a material with higher refractive index, at an angle greater than the critical angle, $\theta_{cr}$ (Fig. 1a-b). In both cases, the electric field just beyond the interface does not abruptly drop to zero, but decays exponentially as an evanescent wave [20]. When the lower-index material has a thickness on the order of the evanescent-wave decay length, light can tunnel through the material and emerge on the other side. This is referred to as frustrated total internal reflection (F-TIR) [21] [22] (Fig. 1c), and has numerous applications such as multi-touch sensing [21]. Analogously, this process can take place for external reflection, which we refer to as *frustrated external reflection* (F-ER; here we omit 'total' from the term due to the non-zero optical losses) (Fig. 1d).

We calculated the decay length of the evanescent wave [23] to approximately obtain the $SiO_2$ film thickness required for observation of F-ER (See the Supplementary Information 3 for further discussion). To demonstrate F-ER experimentally, we deposited three $SiO_2$ films with thicknesses of 1.5, 3, and 6 $\mu m$ onto double-side-polished high-resistivity silicon wafers using PECVD. At $\lambda_0 = 7.8\ \mu m$, these films have $\tilde{n}_{ox} = 0.8 + 0.07i$ (Fig. 2), and thus $\theta_{cr} = 51°$. In Fig. 4a, we show the measured and calculated transmittance versus the incident angle for p polarization. The calculations were performed using the transfer matrix method [24], assuming a silicon substrate thickness of 500 $\mu m$. The fine fringes in the calculated data, which are due to thin-film interference in the silicon, are not observed in the experiment most likely because of the non-zero numerical aperture (*i.e.*, each measurement corresponds to an average over a range of incident angles). The plotted calculated data was smoothed using a moving average to roughly emulate the measurement. Figure 4a shows that some light is transmitted even for $\theta > \theta_{cr}$, demonstrating F-ER. As expected, an increase of the thickness of the $SiO_2$ layer ($d$) results in lower transmittance.

For $\theta > \theta_{cr}$, the reflectance increases with increasing $d$ (Fig. 4b), as the effect of frustration is reduced. For $\theta < \theta_{cr}$, however, increasing $d$ results in a decrease in reflectance. This occurs because $|\tilde{n}_{ox}|$ is close to 1 (*i.e.*, the index of air), so most of the reflected light comes from the SiO$_2$/Si interface rather than the air/SiO$_2$ interface. A thicker SiO$_2$ film absorbs more light, reducing the overall intensity that emerges back into the air. At angles beyond $\theta_{cr}$, the slope of the transmittance vs. angle changes, indicating the onset of F-ER (see Fig. 4a). Note that the reflectance does not reach unity even at very oblique angles because of evanescent-wave absorption within the SiO$_2$ (*i.e.*, no *total* external reflection). See the Supplementary Information 5 for more detailed calculations of a fictitious case with no optical absorption.

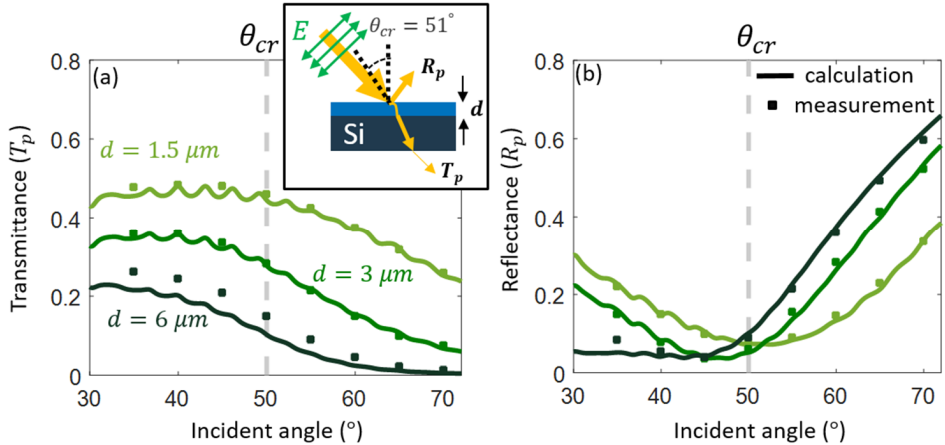

**Figure 4)** Frustration of external reflection. **(a)** Transmittance and **(b)** reflectance of an SiO$_2$ film on a silicon substrate, for p-polarized light versus incident angle at $\lambda_0 = 7.8\ \mu m$, for three different film thicknesses of SiO$_2$, $d$. The $d = 6\ \mu m$ film is significantly thicker than the evanescent decay length $\delta$, so almost no light is transmitted for incident angles beyond $\theta_{cr}$, maximizing reflectance. Conversely, for the thinner films, the transmittance does not drop completely to zero, and the reflectance is reduced, indicating frustrated external reflection (F-ER). $\theta_{cr}$ is identified with a vertical dashed grey line.

### 2.3. Direct coupling to surface plasmon waves from air

The use of materials with $n < 1$ allows us to "invert" many common configurations in optics, using air as a high-index medium (Fig. 1). To demonstrate the generality of this concept, we explored replacing the high-index prism in the well-known Kretschmann configuration for coupling to surface-plasmon polaritons (SPPs) [25] with air. Specifically, we explore coupling to SPPs at the interface between gold (Au) and SiO$_2$, in the 7.4 – 7.7 $\mu$m wavelength region, where $n_{ox} < 1$ and $\kappa < 0.08$ (Fig. 2).

Our simple structure comprises 10 nm of evaporated Au on top of an SiO$_2$ wafer (Fig. 5a). Using the transfer-matrix method, we calculated angle-dependent reflectance from air for slightly different wavelengths (7.45, 7.56, and 7.67 $\mu$m) corresponding to different values of $n_{ox}$ (0.9, 0.84, and 0.76, respectively) (Fig. 5a). The local minima in the reflectance are evidence of coupling to SPPs propagating along the SiO$_2$/Au interface. The positions of these minima shift toward smaller incident angles for increasing wavelength, as expected from the dispersion of SiO$_2$ (Fig. 2a). Figure 5b shows the corresponding experimental results, which are in good agreement with our calculations.

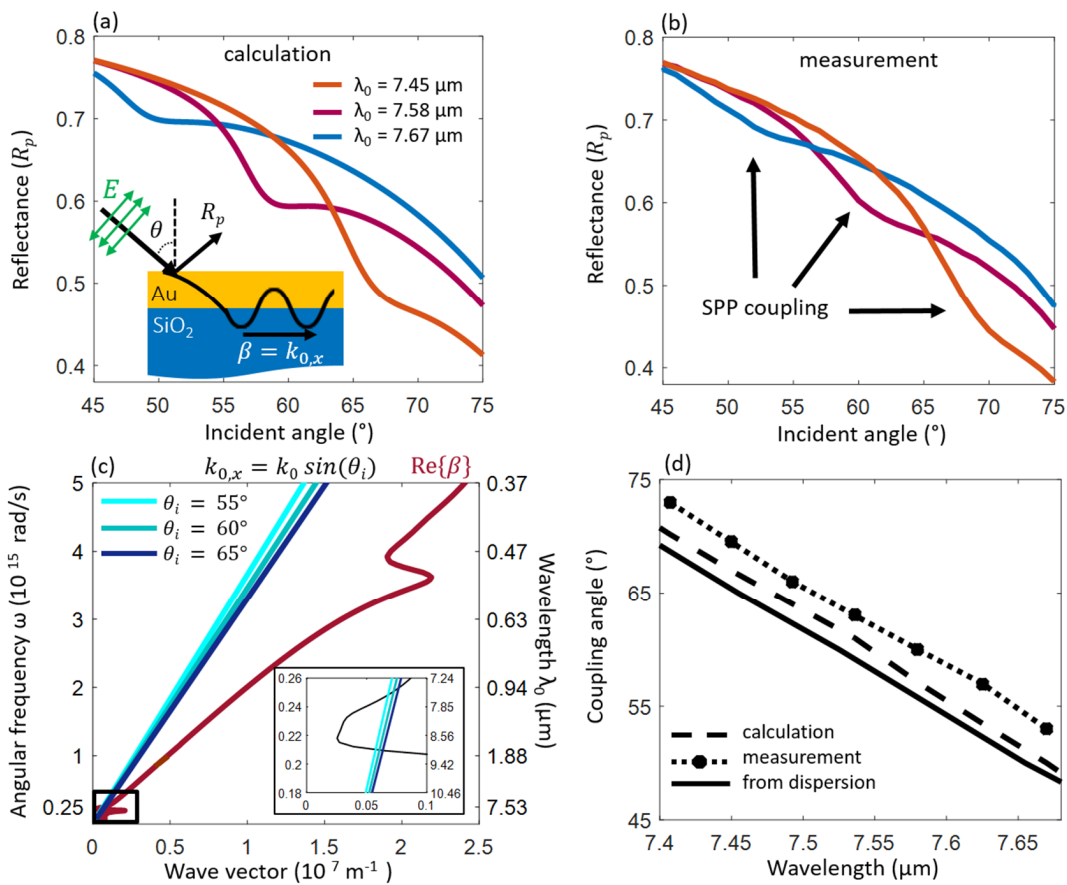

**Figure 5)** Direct coupling to SPPs from free space. **(a)** Calculated and **(b)** measured p-polarized angle-dependent reflectance of a 10 nm Au film deposited on an SiO$_2$ substrate at $\lambda_0 = 7.45, 7.58,$ and $7.67~\mu m$, corresponding to $n_{ox} = 0.9, 0.84,$ and $0.76$, respectively. The dips are evidence of coupling to SPPs at the interface between Au and SiO$_2$. **(c)** Dispersion plots for incident light for different angles, and SPPs at the Au/SiO$_2$ interface. The intersections around 7.5 μm (see inset) correspond to locations of the dips in the reflectance curves in (a, b). Three blue traces show $k_{0,x}$, the x-component of the wave vector of the incident light for $\theta_i = 55, 60, 65$°. Smaller angles of incidence correspond to smaller $k_{0,x}$, and therefore intersections of $k_{0,x}$ and $\beta$ occur at lower frequency or longer wavelength. **d)** Summary of the results in parts (a-b) and (c). The dashed and dotted lines identify the angular positions of the dips in the reflectance curves from (a) and (b). The black solid line corresponds to the intersections between the dispersion curves in (c), signifying coupling to SPPs.

To verify that these minima in reflectance correspond to coupling to SPPs, we plotted the dispersion relation of single-interface SPPs [26] and the dispersion curves of the incident light — taking only the component of the incident wave vector along the interface — for several different incidence angles (Fig. 5c). Direct coupling to SPPs is expected to occur roughly when these curves intersect. It can be seen in the inset of Fig.

5c, which zooms in on the region where $n_{ox} < 1$, that for every angle there are two distinct intersections. Only the shorter-wavelength intersection (~ 7.5 $\mu m$) results in clear coupling to SPPs. The wavelength of the other intersection (~ 9 $\mu m$) corresponds to high losses in the $SiO_2$ (Fig. 2b), so the SPPs are almost immediately absorbed (more information can be found in Supplementary Information 6). In Fig. 5d, we converted the intersections in the dispersion to a plot of the predicted coupling angle versus free-space wavelength (solid line). This curve matches well with the actual measured and calculated coupling angles. The slight discrepancy may be a result of experimental uncertainty and the limitations of the SPP wave-vector equation, which assumes that both the metal and dielectric have semi-infinite thickness [26].

### 2.4. Other materials

The phenomena described in this paper all become easier to observe for materials that simultaneously have $n$ significantly below 1 and minimal losses. However, the frequency dependences of $n$ and $\kappa$ are always connected (as described by the Kramers-Kronig relations [27]) and generally have an inverse relationship; *e.g.*, the reduction of $n$ in $SiO_2$ is accompanied by an increase of $\kappa$ on the lower-loss side of the resonance (Fig. 2). To evaluate candidate materials, we define a figure of merit that simultaneously quantifies the loss and the difference of $n$ from unity: $\eta = \kappa/(1 - n)$. This figure of merit is used in the X-ray regime to quantify the sharpness of the transition between ordinary and external reflection [28] (more discussion in Supplementary Information 7). In Fig. 6, we plotted $\eta$ as a function of frequency in the near and mid infrared for several candidate materials with optical-phonon or plasma resonances [11], [29]. Among the materials we explored, silicon carbide reaches the smallest value of $\eta$ [30], and lithium fluoride has the longest wavelength span for which $n < 1$ and $\eta < 1$ [31]. In general, fluorides seem to be good candidate materials [32], with the wavelength of the $n < 1$ region increasing for larger cations. We note that many of these candidate materials are transparent with $n > 1$ at shorter wavelengths (*e.g.*, in the visible), potentially enabling very different functionalities in different spectral ranges.

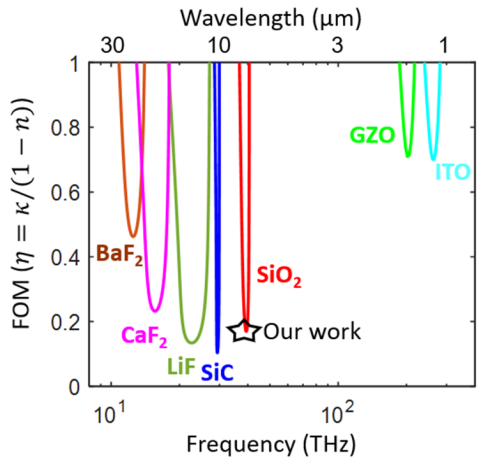

**Figure 6)** Figure of merit, $\eta = \kappa/(1-n)$, for several candidate materials with $n < 1$, including transparent conducting oxides in the near infrared and polar dielectric materials in the mid infrared (material data from refs. [11], [29], [30], [31], and [32].)

3. **Conclusion**

In this article, we explored the use of low-loss materials with refractive index below unity ($n < 1$) to "invert" several well-known configurations in optics that are based on total or attenuated internal reflection. We showed that polar dielectrics with strong phonon resonances are good candidate materials for such applications, experimentally demonstrating *external reflection* in the mid-infrared spectral range, *frustrated external reflection*, and *direct coupling to surface plasmon polaritons* from free space without using prisms or surface modulation. We envision potential applications of the $n < 1$ region in polar dielectrics and other materials with low-loss resonances for spectral or angular filtering, air-core waveguiding, and surface-plasmon-based sensing.

**Acknowledgements**

We acknowledge funding from the Office of Naval Research (Grant No. N00014-16-1-2556) and UW Madison. BS acknowledges a Graduate Research Fellowship from the National Science Foundation. PJR acknowledges a Critical Skills Master's Fellowship from Sandia National Labs. We thank M. Shahabadi and R. Wambold for helpful discussions. The PECVD of SiO$_2$ was performed at the Harvard Center for Nanoscale Systems by P. de Rouffignac. Various characterization was performed at the Materials Science Center and Soft Materials Laboratory, core facilities at UW-Madison. We acknowledge use of facilities and instrumentation supported by NSF through the University of Wisconsin Materials Research Science and Engineering Center (DMR-1720415).

Supplementary information for
**Mid-infrared optics using dielectrics with refractive indices below unity**

Alireza Shahsafi[1], Yuzhe Xiao[1], Jad Salman[1], Bradley S. Gundlach[1], Chenghao Wan[1, 2], Patrick J. Roney[1], Mikhail A. Kats[1, 2, 3]

Departments of [1]Electrical and Computer Engineering, [2]Materials Science and Engineering, and [3]Physics

University of Wisconsin-Madison

## 1. Measurement of $n_{ox}$ and $\kappa_{ox}$ for SiO$_2$

For the bulk SiO$_2$ samples used in Section 5, optical properties were extracted using variable-angle spectroscopic ellipsometry (J. A. Woollam, Co. IR-VASE). The data was fit using a model comprising two Gaussian oscillators for the complex permittivity ($\epsilon = \epsilon_1 + i\epsilon_2$):

$$\epsilon_1 = \Gamma\left(\frac{E-E_n}{B_r}\right) + \Gamma\left(\frac{E+E_n}{B_r}\right), \quad (S.1)$$

$$\epsilon_2 = Ae^{-\left(\frac{(E-E_n)}{B_r}\right)^2} + Ae^{-\left(\frac{(E+E_n)}{B_r}\right)^2} \quad (S.2)$$

where A is the amplitude, E$_n$ is the center energy, and B$_r$ is the broadening of the oscillator. The function $\Gamma$ produces a Kramers-Kronig-consistent line shape, and is defined in ref. [S1]. The resulting fit parameters are shown in Table 1.

However, we were unable to use ellipsometry to extract the optical properties of our PECVD-deposited SiO$_2$ samples due to the need to preserve the polished backside surface for other measurements. Without roughening the backside of the sample, interference effects from that interface can substantially reduce the accuracy of ellipsometric modeling [S1]. Instead, we determined the optical constants by reverse calculating the values from reflectance and transmittance measurements. We measured the angle- and wavelength-dependent reflectance and transmittance of our thinnest SiO$_2$ films (1.5 $\mu$m) on silicon, over the 7 – 9 $\mu$m range. We then used the transfer-matrix method [S2] to calculate the corresponding reflectance and transmittance for an unknown film defined by a complex refractive index that we varied across the range of reported refractive index values for SiO$_2$ [S3-4]. To determine which of these best corresponded to our films, we minimized the function $S(\lambda)$ for each wavelength $\lambda$, defined as:

$$S(\lambda) = \sum_{\theta,s,p}(\delta_{(s,p),\theta} + \Delta_{(s,p),\theta}) \quad (S.3)$$

$$\delta_{(s,p),\theta} \triangleq |R_{m,(s,p),\theta} - R_{bf,(s,p),\theta}|$$

$$\Delta_{(s,p),\theta} \triangleq |T_{m,(s,p),\theta} - T_{bf,(s,p),\theta}|$$

where $T_m$ and $R_m$ are the measured transmittance and reflectance, $T_{bf}$ and $R_{bf}$ are the corresponding calculated values, s/p identifies the polarization, and $\theta = 35°, 40°,$ and $45°$ are the incident angles used. The resulting "brute-force-calculated" values of $n_{ox}$ and $\kappa_{ox}$ were then fit to the same Gaussian oscillator model, as shown in Table 1 and Fig. S1. The optical properties of our PECVD SiO$_2$ extracted using this method are also shown in Fig. 2 (dotted lines). As can be seen in Fig. 4, the calculated reflectance and transmittance using the extracted optical properties for SiO$_2$ matched reasonably well with the measurement even for incidence angles greater than 45° or other film thicknesses (3 or 6 $\mu m$ SiO$_2$).

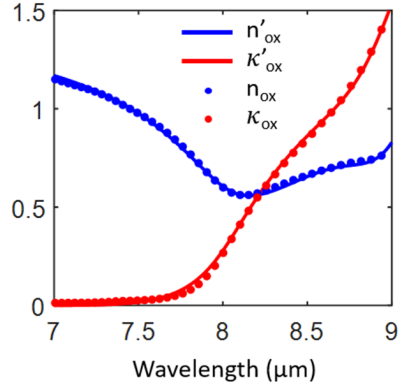

**Figure S1)** Optical properties of PECVD SiO$_2$ from fitting to a Gaussian model ($n'_{ox}$, $\kappa'_{ox}$) and from a wavelength-by-wavelength fit of the transmittance and reflectance data ($n_{ox}$, $\kappa_{ox}$)

Table 1. Gaussian oscillator model properties for SiO$_2$ optical properties (see also the plots in Fig. 2)

|  | Amplitude, A | | Center energy, $E_n$ (cm$^{-1}$) | | Broadening, $B_r$ (cm$^{-1}$) | |
| --- | --- | --- | --- | --- | --- | --- |
| Bulk SiO$_2$ | 10.55 | 1.01 | 1070.4 | 1092.2 | 63.58 | 228.63 |
| PECVD SiO$_2$ | 7.92 | 1.65 | 1037.5 | 1105.5 | 81.66 | 193.2 |

## 2. Pseudo-Brewster angle

The Brewster angle is the angle of incidence for which there is no reflection of p-polarized light from an interface between two lossless transparent materials [S5]. When light is incident onto a material with optical loss, such an angle does not exist, but there is still an angle for which p-polarized reflection is minimized, such that $\frac{d|r_p^2(\theta_{pB})|}{dt} = 0$; this is known as the pseudo-Brewster angle ($\theta_{pB}$) [S6]. $\theta_{pB}$ is shown in Fig. S2a for the case of light coming from air ($n_1 = 1$) and incident on an absorbing material described by some $n = \text{Re}\{n_2\}$ and $\kappa = \text{Im}\{n_2\}$. This figure shows that for large $n_2$ (e.g., $n$ and $\kappa$ greater than 5), $\theta_{pB}$ is large (*e.g.*, greater than 60°). For the case of a noble metal in the mid infrared range, $\theta_{pB}$ is greater than 85°. The high-angle p-polarized reflectance of a gold mirror is plotted in Fig. S2-b for the mid-infrared wavelength range, showing a decrease of reflectance because of the pseudo-Brewster effect. The spectral and angular window in which SiO$_2$ has higher reflectance than gold is identified with a black box. The inset in this plot shows that p-polarized reflectance of SiO$_2$ surpasses the reflectance of Au at $\lambda = 7.4\ \mu m$ starting at an incident angle of 85°. This calculation was done using the transfer matrix method [S2] and literature optical properties for gold [S7].

**Figure S2)** (a) Calculated pseudo-Brewster angle ($\theta_{pB}$) for light incident from air onto a material with some arbitrary $n$

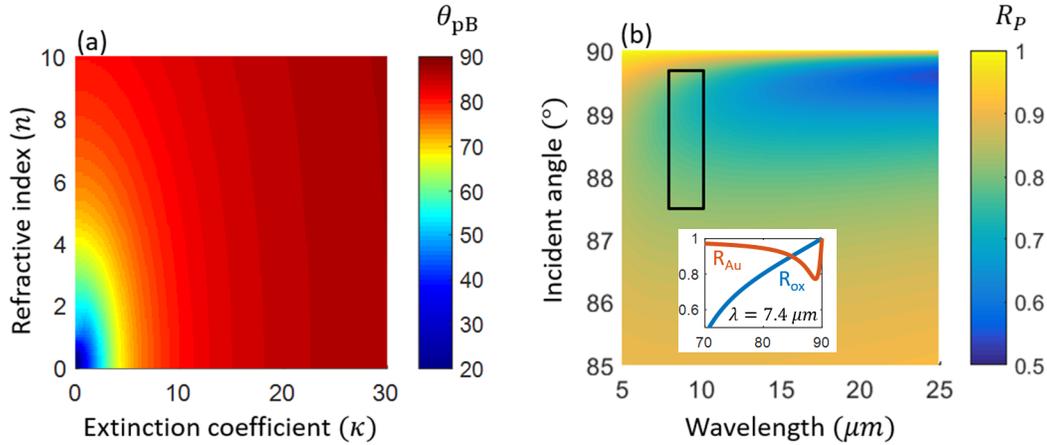

and $\kappa$. (b) In the wavelength range in which the refractive index of SiO$_2$ is less than unity, the p-polarized reflectance of a gold mirror can decrease down to 75% because of the pseudo-Brewster effect.

## 3. Decay length calculation for SiO$_2$

Based on a calculation of the decay length of the evanescent wave inside the SiO$_2$ film, we selected three thicknesses to deposit to experimentally demonstrate frustration of external reflection. The decay length ($\delta$) is defined as the distance where the electric field amplitude of an incident wave of light decays to $1/e$ of its value just after the interface. Assuming a plane wave, the z-dependence of the electric field is given as $E_0 e^{ik_z z}$, where $k_z$ is the wavevector of light in the z direction, which is normal to SiO$_2$ surface. The analytical expression for the decay length of the electric field is [S8]:

$$\delta(\lambda_0, \theta) = \frac{1}{imag(k_z)} = \frac{1}{\text{Im}(\tilde{n}_{ox} k_0 \cos\theta_2)} = \frac{\lambda_0}{2\pi} \frac{1}{\text{Im}\left(\tilde{n}_{ox}\sqrt{1 - \frac{n_{air}^2}{\tilde{n}_{ox}^2}\sin^2\theta}\right)} = \frac{\lambda_0}{2\pi} \frac{1}{\text{Im}(\sqrt{\tilde{n}_{ox}^2 - \sin^2\theta})} \quad (S.4)$$

in which $\tilde{n}_{ox}$ is complex refractive index of SiO$_2$, $\lambda_0$ is wavelength and $\theta_2$ and $\theta$ are the angle in SiO$_2$ and in air, respectively. By using the optical properties of PECVD SiO$_2$, $\delta$ is calculated and shown in Fig. S3. For incident angles greater than the critical angle (shown by the grey line), $\delta$ decreases rapidly (*i.e.*, external reflection occurs).

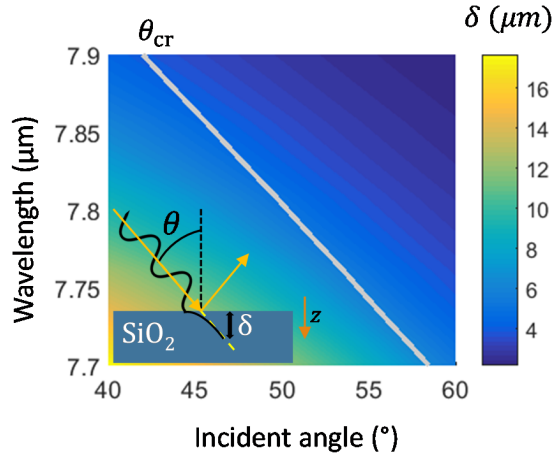

**Figure S3)** The decay length, $\delta$, as a function of wavelength and incident angle, for light incident from air onto an SiO$_2$ slab. For larger wavelengths and/or angles, the decay length decreases dramatically. The gray line on top of the color plot shows the critical angle

**5. Frustration of external reflection for lossless SiO$_2$**

Fig. S4 shows the calculation of reflection and transmission versus incident angle for the same structure of SiO$_2$ on top of silicon substrate, assuming there is no optical loss in the SiO$_2$ ($\kappa_{ox} = 0$), but the same $n_{ox}$ as in the main text. This calculation was done to analyze frustrated *total* external reflection, and it shows that the transition to total reflection occurs much faster across the critical angle compared to the real case with optical loss. Similarly to the real case with losses, the transition happens more gradually for thinner films. Note that there exists an angle in both the angular reflection and transmission plots at which the curves for different thickness overlap; this corresponds to the Brewster angle at the SiO$_2$/Si interface. At this angle, the reflection from the structure is dominated by the air/SiO$_2$ interface; therefore, reflection becomes independent of the film thickness.

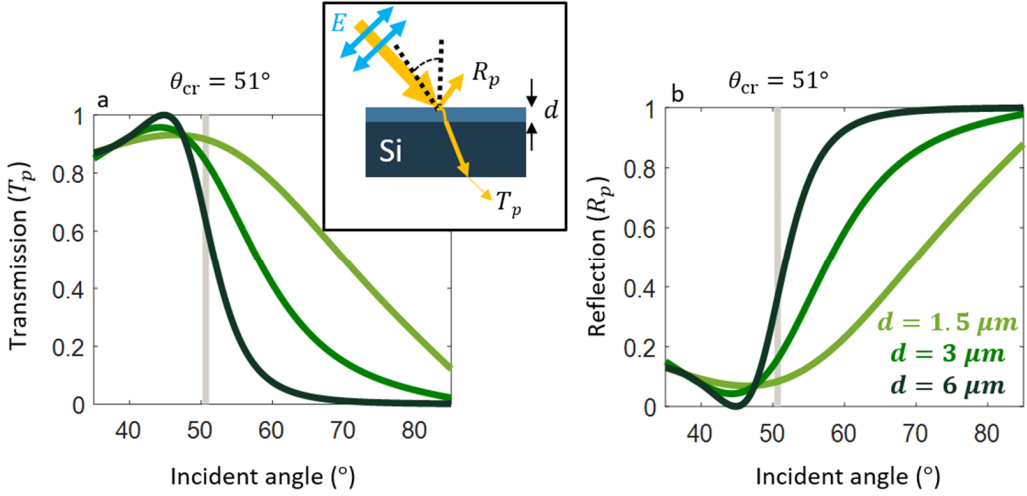

**Figure S4)** (a) Transmittance and (b) reflectance of a fictitious $SiO_2$ film without optical loss on a silicon substrate, for p-polarized light, versus incident angle at $\lambda_0 = 7.8\ \mu m$, for three different film thicknesses of $SiO_2$, $d$. $\theta_{cr}$ is identified with a vertical grey line.

## 6. Other intersections in the surface-plasmon dispersion plot

In this section, we plot the real and imaginary parts of the propagation constant $\beta$ of single-interface surface plasmon polaritons (SPPs) [S9] together with the dispersion curves of the incident light from free space — taking only the component of the incident wave vector along the interface — for several different incidence angles (Fig. S5). The inset of Fig. S5 is a zoom-in of the region where $n_{ox} < 1$ and it shows two spectral regions in which Re$\{\beta\}$ crosses the dispersion curves of incident light. In the main manuscript, we did not focus on the crossing points that happens at larger wavelength (around $9\ \mu m$) because, as is shown here, Im$\{\beta\}$ is very large, and thus SPPs are immediately absorbed.

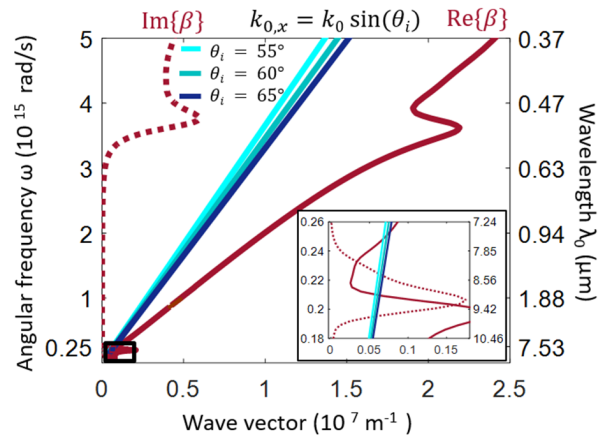

**Figure S5)** Dispersion plot for incident light for different angles, and SPPs at the gold/SiO$_2$ interface. Here, we also plot Im{$\beta$} which is a measure of the intrinsic loss of SPPs.

## 7. Transition to total external reflection

In this section, the reflection of light at an interface between air and a material with $n < 1$ for different polarizations of light versus incident angle is shown. In Fig. S6(a, b), it is assumed that $n$ is constant and smaller than one (in this case, 0.7), and $\kappa$ varies. These plots show that a sharper transition to external reflection can be achieved by using a material with lower loss.

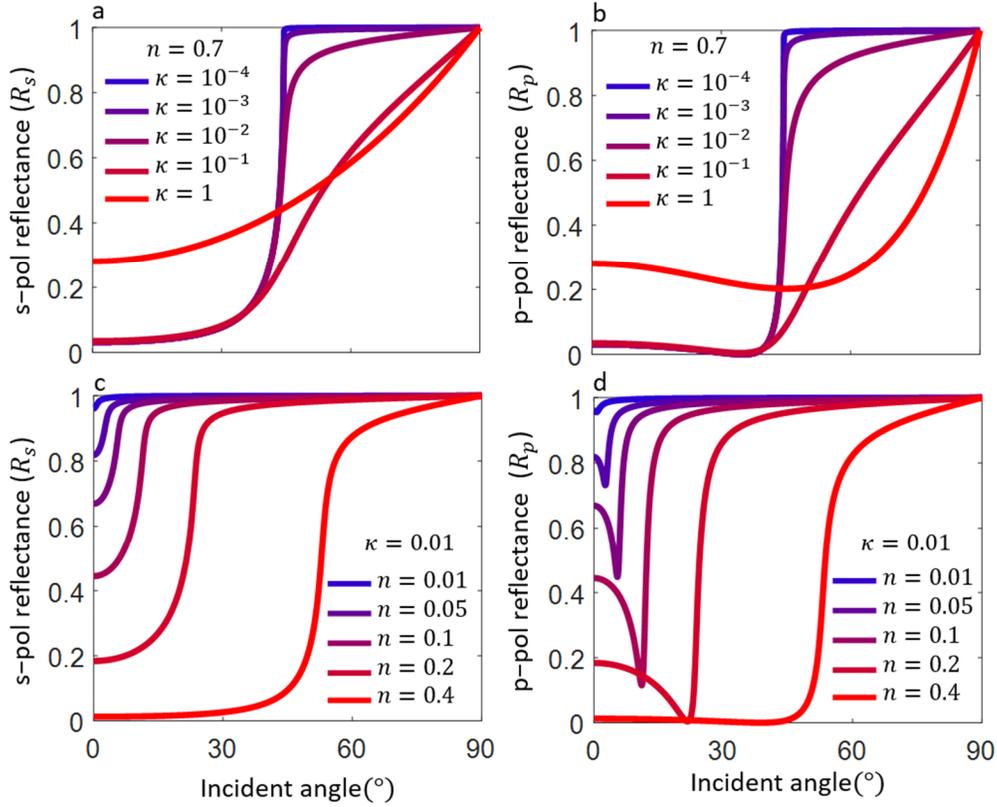

**Figure S6) Transition to total external reflection with various values of $n$ and $\kappa$**
(a,b) The increase of optical loss ($\kappa$) smooths the transition to total reflection. (c, d) Change of refractive index ($n$) changes the critical angle.

Setting a constant $\kappa$ for the material with $n < 1$ and sweeping $n$ results in different critical angles of external reflection (Fig. S6(c, d)). Also, increasing $n$ results in a more gradual transition. To consider the effect of both $n$ and $\kappa$, we used a figure of merit, $\eta = \kappa/(1 - n)$, used at X-ray frequencies [S10]. A decrease of $\eta$ results in a sharper transition to total external reflection. In Fig. S7, we plot the polarization-averaged reflection of several materials from Fig. 6, versus the incident angle divided by the critical angle, at the wavelengths at which $\eta$ is minimum.

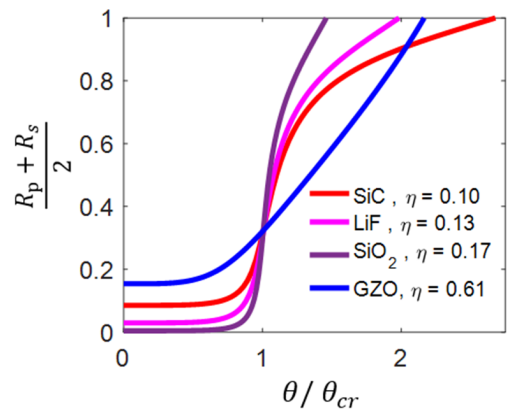

**Figure S7)** Polarization-averaged reflectance versus incident angle, normalized to the critical angle, for several representative materials from Fig. 6. Each plot is at the wavelength for which $\eta$ is minimum for that material.